\documentclass[aps,preprint,a4paper,showpacs,amsmath,amssymb,floatfix]{revtex4}
\usepackage{graphicx}% Include figure files
\usepackage{dcolumn}% Align table columns on decimal point
\usepackage{bm}
\newcommand{\Op}[1]{{{\mathrm{\hat{#1}}}}}

\begin{document}
%=================================================================
% Full title of the paper (Capitalized)
\title{Quantum thermodynamics  in strong coupling: heat transport and refrigeration.}

% Authors (Add full first names)
\author{Gil Katz $^{1,\ddagger}$ and Ronnie Kosloff $^{2,\ddagger}$ }
%\AuthorNames{} %%MDPI internal note: new layout%%

% Affiliations / Addresses (Add [1] after \address if there is only one affiliation.)
\affiliation{%
$^{1}$ \quad Institute  of Chemistry  The Hebrew University, Jerusalem 91904 Israel and Afikim Science, Israel; gil@afimilk.com\\
$^{2}$ \quad Institute  of Chemistry  The Hebrew University, Jerusalem 91904, Israel; ronnie@fh.huji.ac.il}

% Abstract (Do not use inserted blank lines, i.e. \\) 
\begin{abstract}
The performance characteristics of a heat rectifier and a heat pump are studied in a non Markovian framework. 
The  device is constructed from a molecule connected to a hot and cold reservoir.
The heat baths are modelled using the stochastic surrogate Hamiltonian method. The molecule  
is modelled  by an asymmetric double-well potential. Each well is semi-locally connected to a heat bath composed of spins.
The dynamics is driven by a combined system-bath Hamiltonian.
The temperature of the baths is regulated by a secondary spin bath composed of identical spins in 
thermal equilibrium. A random swap operation exchange spins between the primary and secondary baths.
The combined system is studied  in various system-bath coupling strengths. 
In all cases the average heat current always flows from the hot towards  the cold bath
in accordance to the second law of thermodynamics.
The asymmetry of the double well  generates a rectifying effect meaning that when the left and right baths are exchnaged 
the heat current follows the hot to cold direction.
The heat current is larger when the high frequency is coupled to the hot bath. 
Adding an external driving field can reverse the transport direction. 
Such a refrigeration effect is modelled by  a periodic driving field in resonance with the
frequency difference of the two potential wells. A minimal driving amplitude is required to overcome the
heat leak effect. In the strong driving regime the cooling power is non-monotonic with the system-bath coupling.
\end{abstract}

\maketitle

\section{Introduction}

Quantum thermodynamics can address a single microscopic quantum device performing thermodynamical tasks
such as heat transport, power generation and refrigeration. A dynamical perspective is incorporated in the field of open quantum systems \cite{k281}. 
The established approach is based on the weak coupling system-bath approximation \cite{davies1974markovian} leading to
Markovian dynamics described by the L-GKS master equation \cite{lindblad1976generators,gorini1976completely}.
This approach leads to dynamical evolution which is  consistent with the laws of thermodynamics \cite{k281,alicki1979quantum}.
In the present paper we deviate from the established scenario and study a quantum device strongly coupled to the heat baths
and undergoing non Markovian dynamics. 

First we study the phenomena of heat transport. The second law of thermodynamics
dictates that heat should always flow from the hot to the cold bath. We use this phenomenon to benchmark our model.
Once established we add to the model an external driving field. Then the device can operate as a refrigerator pumping heat
from the cold bath to the hot bath.

The phenomenon of transport is associated with a gradient of a driving potential. Even a single molecule subject to
temperature gradient will experience a heat current flowing from the high potential to the lower one. 
Modelling such a heat flow is the subject of this study.
A novel idea is heat rectification, the possibility of asymmetric heat transfer when the baths are switched
\cite{segal2003thermal,segal2005spin}. Employing such a concept one can imagine effectively isolating a subsystem 
in a far from equilibrium state. Close to equilibrium  the rectifying effect must vanish.
In a series of papers Segal and co workers studied the conditions for 
heat rectification tracing the effect to non linear phenomena either related to asymmetric baths (case A) or to asymmetric 
system bath coupling (case B) \cite{wu2009sufficient,segal2009absence}.  In the present paper  heat transport of case B is analysed
based on the non Markovian stochastic surrogate hamiltonian approach \cite{k238,k259}. 

The popular  theoretical approach to describe heat transport in a network is to compose the transport equations 
from local equations. This local approach is correct when only populations are involved.
However, when quantum systems are involved the local approach leads to the violation of the second law of Thermodynamis \cite{k290}. 
A thermodynamically consistent treatment within the weak coupling system-bath approximation requires first to diagonalize the total subsystem network  obtaining the global  transition frequencies of the total system \cite{k281}.  An alternative is a careful perturbation analysis \cite{trushechkin2015perturbative}.
Only then the weak coupling procedure can be applied to obtain the consistent  transport equations.  
A major assumption in the derivation is a tensor product form of the system and baths at all times.
The weak coupling procedure employed in Ref. \cite{wu2009sufficient,segal2009absence} is global and consistent with thermodynamics. 

The standard treatment of transport is based on the  Landauer-Buttiker scattering theory \cite{buttiker1985generalized}. 
Each scattering event which transports energy from one bath to the other is assumed to be independent 
resulting in a Poissonian process. This ballistic description ignores intricate dynamics between
the molecular device and its leads.

Can one derive a consistent transport theory beyond the weak system bath coupling limit? An attempt in this direction
has been described by Segal \cite{segal2014two,jing2015transient} who studied a two-level-system coupled to a  spin bath
in the intermediate coupling limit.  The derivation is based on the NIBA approach which allows to evaluate the memory kernel explicitly. 
The study finds that the rectifying effect exists in the intermediate coupling range and vanishes 
in the weak coupling limit of the two-level-system  to the spin baths. The NIBA method can only be employed for very simple	
system Hamiltonians such as a two-level-system or harmonic oscillators. 
This approach using the  polaron transformation was employed to constuct a model of solar energy harvesting \cite{xu2016polaron}.
Therefore a more general approach appropriate for an arbitrary system is required.
Recently the  Nonequilibrium Green's Function Approach (NGFA) has  been suggested as a framework of quantum thermodynamics \cite{esposito2015}.
In this approach the energy of the system and its coupling to the reservoirs are controlled by a slow
external time-dependent force treated to first order beyond the quasistatic limit.

In the present study of heat transport a correlated system-bath scenario is employed.
We  defy the assumption used in all previous studies  of an uncorrelated system and baths represented as a tensor product   
$\hat \rho = \hat \rho_L \otimes \hat \rho_s \otimes {\hat \rho}_R$ \cite{breuer2002theory}.
In addition the system Hamiltonian is represented on a grid, practically achieving
a Hilbert space of the size of $10^3$. The transport dynamics is simulated 
based on the stochastic surrogate hamiltonian approach \cite{k238,k259}. 
The method represents the state by a global system-bath wave function.
This leads to a strongly entangled state at all times. As a result the dynamics is non Markovian. A stochastic layer 
is employed to trim the calculation and define the bath temperature. We check the validity of the approach by
evaluating that the total entropy generation is positive. Once the method is established the asymmetric rectifying effect with identical spin baths is studied.

By adding an external driving field the device is transformed to a refrigerator, driving heat from the cold to the hot bath.
This device falls in the category of continuous quantum engines \cite{k289}.
Optimal performance will be achieved when the driving frequency is in resonance with the frequency difference of the left and right bath.
The refrigeration is studied in the strong coupling regime. It is expected that the cooling current deviates from a linear
relation with the coupling. Such an effect has been observed using the NIBA approach \cite{gelbwaser2015strongly}.

\section{Basic construction}

In the stochastic surrogate hamiltonian approach
the molecular device is subject to dissipative forces due to coupling to two primary baths. 
In turn, the primary baths are subject  to interactions with a secondary baths:

\begin{equation}
 \Op H_{T}=\Op H_S~+~\Op H_{BR}+\Op H_{BL} +\Op H_{B^{``}R}+\Op H_{B^{``}L} ~+~ \Op H_{SBR} ~+~ \Op H_{SBL} + \Op H_{BB^{``}R}~ +~ \Op H_{BB^{``}L} ,
\label{eq:ham}
\end{equation}
where $\Op H_S $ represents the system, $\Op H_{BR}$ and  $\Op H_{BL}$ represents the primary baths,
$\Op H_{B^{``}R}$ and $\Op H_{B^{``}L}$ the secondary bath's, $\Op H_{SBR}$ and $\Op H_{SBL}$ the system-bath interaction.
$\Op H_{BBR^{``}}$ and $\Op H_{BBR^{``}}$ are the primary/secondary bath's interactions. 
The system Hamiltonian $ \Op H_S$ describes  molecular nuclear modes:
The molecular Hamiltonian  has the form:
\begin{equation}
\Op H_S = \frac{1}{2m} \Op P^2 +V(\Op R)
\end{equation}
Specifically the  molecular device under study is composed of the lower electronic surface of two coupled diabetic  oscillators:
\begin{equation}
V(\Op R) ~=~ \frac{1}{2}\left(V_L(\Op R)+V_R(\Op R) -\sqrt{ |V_L(\Op R)-V_R(\Op R)|^2-4 V_C(\Op R)^2~}\right)
\end{equation}
where:
\begin{equation}
V_{R/L} (\Op R)= \frac{ m \omega_{R/L}^2}{2} (\Op R-R_{R/L})^2
\end{equation}
where parameters are described in Table 1.

The non adiabatic coupling potential: 
\begin{equation}
V_C( \Op R )= A \exp ( -(\Op R-R_0)^2/2 \sigma^2)
\end{equation}
where $\sigma=0.5$, $A=0.5$ and $R_0$ is the crossing point.

The bath is described by a fully quantum formulation.
Briefly, the bath is divided into a primary part interacting with the system directly and a secondary bath
which eliminates recurrence and imposes thermal boundary conditions.

The primary bath Hamiltonian is composed of a collection of two-level-systems.
\begin{equation}
\Op H_{B(R/L)}~~=~~ \sum_j \omega_j \hat \sigma^+_j \hat  \sigma_j ^-
\label{eq:bathham}
\end{equation}
The energies $\omega_j$ represent the spectrum of the bath. 
$\omega_{min}=4.5 \cdot 10^{-5}~a.u.$ and $\omega_{max}=2. \cdot 10^{-3}~a.u.$.

The system bath coupling Hamiltonian has the form:
\begin{equation}
\Op H_{SB(R/L)}=f_{(R/L)}(\Op R) \otimes \sum_j \lambda_j (\hat \sigma^+_j + \hat  \sigma_j^-) 
\end{equation}
$\lambda=0.2~a.u.$ $\lambda_j=\lambda/(\omega_j-\omega_{j-1})$.
The system bath coupling function is chosen to be localised exponentially on the right or left potential well
\begin{equation}
f_{R/L}(\Op R) =  \Gamma \exp ( -\gamma|\Op R - R_{R/L}| )
\end{equation}
where $\gamma=0.5~a.u.$, $\Gamma=0.5~bohr$

The global state of the system and the hot and cold primary baths are realized by a combined wavefucntion.
The dimension of this wavefucntion is of the order of $10^7$. 

The left and right secondary baths are constructed from  virtual spins which are identical to the spin manifold 
of the right and left baths. The state of this bath is a tensor product of spin wavefunctions
with thermal amplitude and random phase \cite{k238}:
\begin{eqnarray}
| \Psi_{B"} \rangle  =\prod_j \otimes |\phi_j \rangle,\\
\nonumber
|\phi_j \rangle =\frac{1}{\sqrt{Z}} \left(
\begin{array}{c}
e^{\frac{\hbar \omega_j}{4 k_B T}+i\theta_1}\\
e^{\frac{-\hbar \omega_j}{4 k_B T}+i\theta_2}
\end{array}
\right)
\end{eqnarray}
where $Z$ is the partition function $Z= 2 \cosh[ \frac{ \hbar \omega_j}{k_B T}]$, $\omega_j$ is the frequency of the $j$th spin
and $\theta_1, \theta_2$ are random phases. The temperature $T$ corresponds to either $T_L$ or $T_R$.
Random swap operations are employed to switch between the wavefuctions of the primary and secondary baths \cite{k293}.
As a result, the secondary bath defines the temperature of the left or right contacts with the system.
Only spins on resonance are swapped therefore the transfer of energy can be defined as heat \cite{barra2015thermodynamic}.

The energy balance relation is equivalent to the first law of thermodynamics.
The time derivative of the system energy defines two heat currents:
\begin{equation} 
 \frac{d E_S}{dt} = \langle i[\Op H_{SBR},\Op H_{S}]+i[\Op H_{SBL}, \Op H_{S}] \rangle ={\cal J}_R+{\cal J}_L
\label{eq:flux}
\end{equation}   
This leads to: 
\begin{equation}
{\cal J}_{R/L}= \frac{1}{2m}\left \langle \left(\Op P \frac{d }{dR} f_{R/L}(\Op R)+\frac{d }{dR} f_{R/L}(\Op R)\Op P \right) \otimes
 \sum_j \lambda_j (\hat \sigma^+_j + \hat  \sigma_j^-) \right \rangle
 \label{eq:flux}
\end{equation}
Comment: if the system-bath coupling was a $\delta$ function, Eq. (\ref{eq:flux}) would become the flux operator.

Figure 1 shows a schematic view of the systems and primary and secondary baths.
\begin{figure}[tb]
\vspace{2.2cm} 
\center{\includegraphics[height=12.5cm]{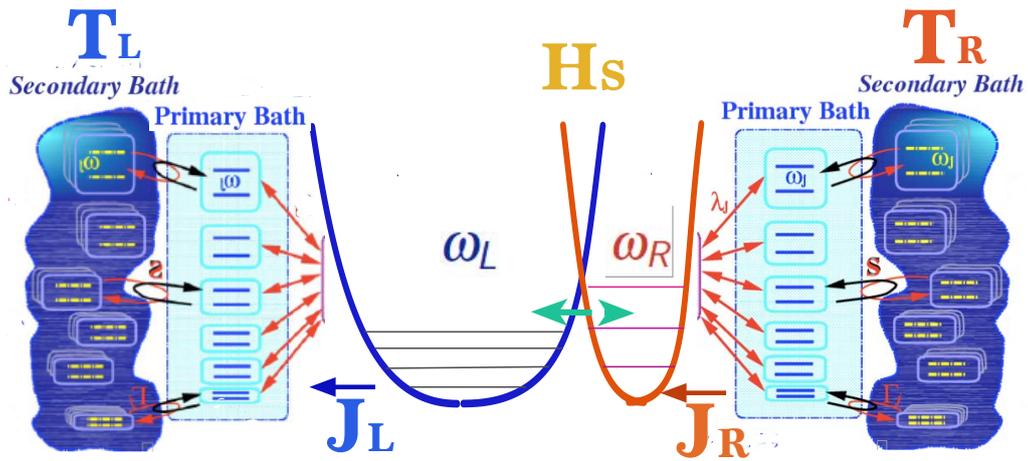}} 
\vspace{0.2cm} 
\caption{ The heat transport setup. In the centre is the system composed of
an asymmetric double well with harmonic frequencies $\omega_L$ and $\omega_R$.
Two primary baths are locally coupled to the left and right wells. Secondary baths with 
identical frequencies as the primary baths impose a temperature by random swap operations.
} 
\label{fig:1}   
\end{figure}
The initial state is the vibrational ground state of the potential ${\Op V}_g(r)$ obtained by propagation in imaginary time \cite{k42}.
The system wavefunction is represented on a Fourier grid of $N_r$ points \cite{k56}.  The combined computation
effort of evaluating the Hamiltonian scales as $N \log N$ where $N$ is the total size of the Hilbert space which includes system grid points and
the dimension of the two baths  \cite{k135}.
Propagation is carried out by the Chebychev expansion of the evolution operator \cite{k28}. The computational effort
scales semi-linearly with $N$ for each stochastic realization. The  number of realizations to achieve convergence is small
and scales as $O(\sqrt{N})$ \cite{k238}. In the present study ten realizations were sufficient.\\ 
Table I summarises the computation details.

\section{Thermodynamical aspects of the Surrogate Hamiltonian}

The stochastic surrogate Hamiltonian approach relies on a large wavefunction to describe the state
of the system and hot and cold primary baths. The typical Hilbert space size is $N \sim10^{7}$. 
The thermalization properties of the SSH have been studied \cite{k238}.
With this size of Hilbert space on the order of 10 realizations are sufficient to converge the local expectation values
of system energy and heat currents. The fast convergences can be related to
quantum typicality meaning that the vast majority of all pure states featuring a common expectation 
value of some generic observable at a given time will yield very similar expectation values of 
the same observable at any later time \cite{bartsch2009dynamical,cramer2010quantum,jin2016eigenstate}. 

The temperature of the hot and cold baths is imposed by the secondary baths.
The spin of the primary bath is stochastically swapped with an  identical spin of the secondary bath.
The result is pure heat transfer between the baths leading to equilibration.
In addition the swap operation generates dephasing of the system \cite{torrontegui2015}.

\subsection{Transport dynamics.}
\label{sec:trans}

\begin{figure}[tb]
\vspace{2.2cm} 
\center{\includegraphics[height=7.5cm]{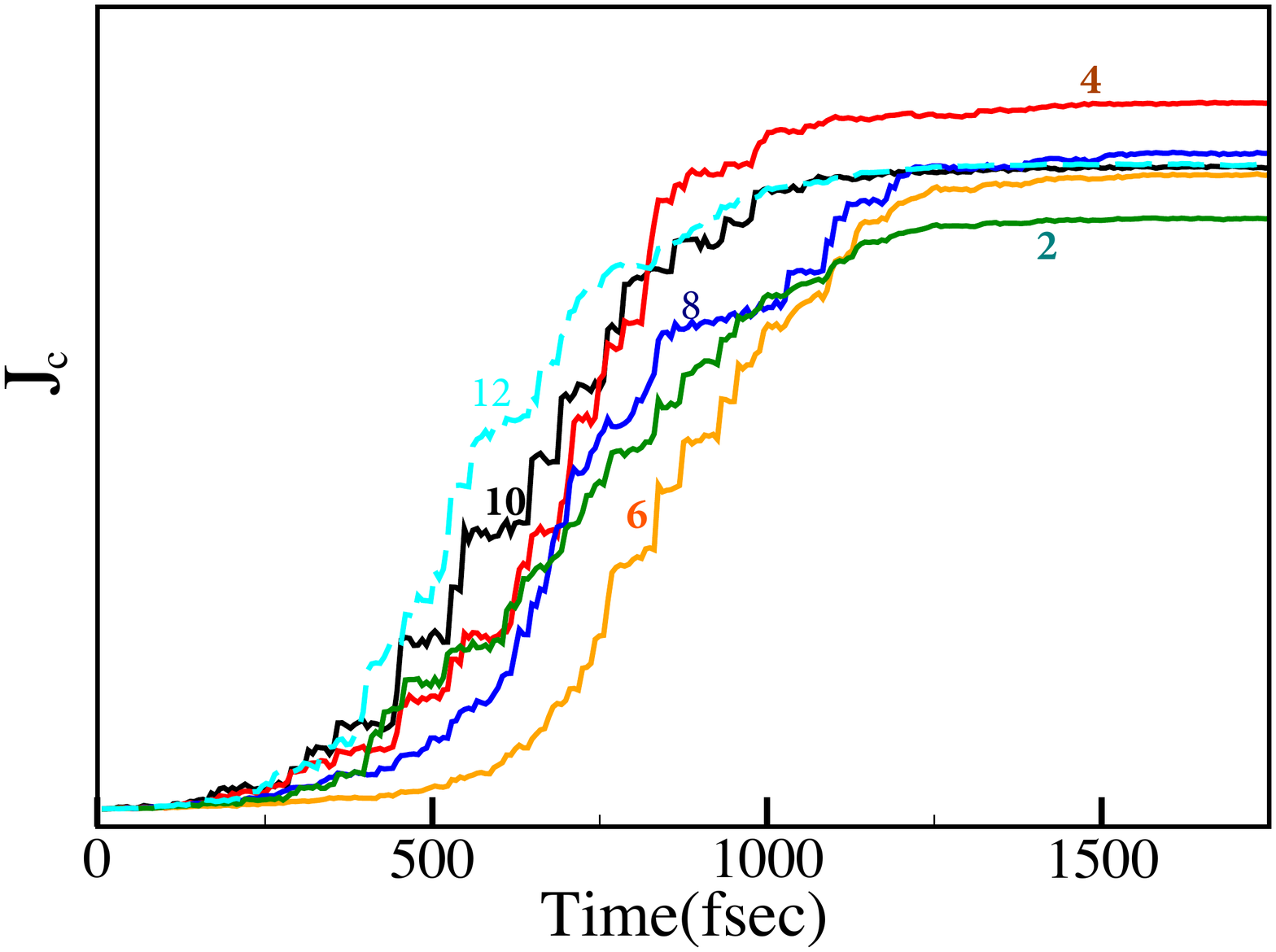}} 
\vspace{0.2cm} 
\caption{The convergence of the current as a function of the number of realizations.
$J_c$ is shown for 2, 4, 6, 8, 10 and 12 realizations. $\omega_L=0.1$, $\omega_R=0.2$ $T_L=10$K, $T_R=25$K.
Number of bath modes is 8 in each bath. The fluctuations become smaller when the steady state is approached.
} 
\label{fig:33}   
\end{figure}

To test the thermodynamical validity of the construction we study heat transport.
Irrespective of the details of the system heat should flow from the hot to cold reservoir.
At first the molecular model is coupled to the right and left thermal leads.
A typical simulation  is initiated from an arbitrary  wavefunction of the system-bath product state. 
The dynamics is then switched on for a sufficient propagation period
until the combined system reaches a steady state.
Each individual realizations shows significant fluctuations.
Averaging even 10 realizations is sufficient to obtain a smooth average Cf. Fig. \ref{fig:33}.
Heat currents are calculated in steady state, then ${\cal J}_L={\cal J}_R$.
The convergence with respect to the number of realizations is shown in Fig. \ref{fig:33}. The steady state current converges 
within 10 realizations. The transient current requires a much larger number of realizations to converge.

\begin{figure}[tb]
\vspace{2.2cm} 
\center{\includegraphics[height=7.5cm]{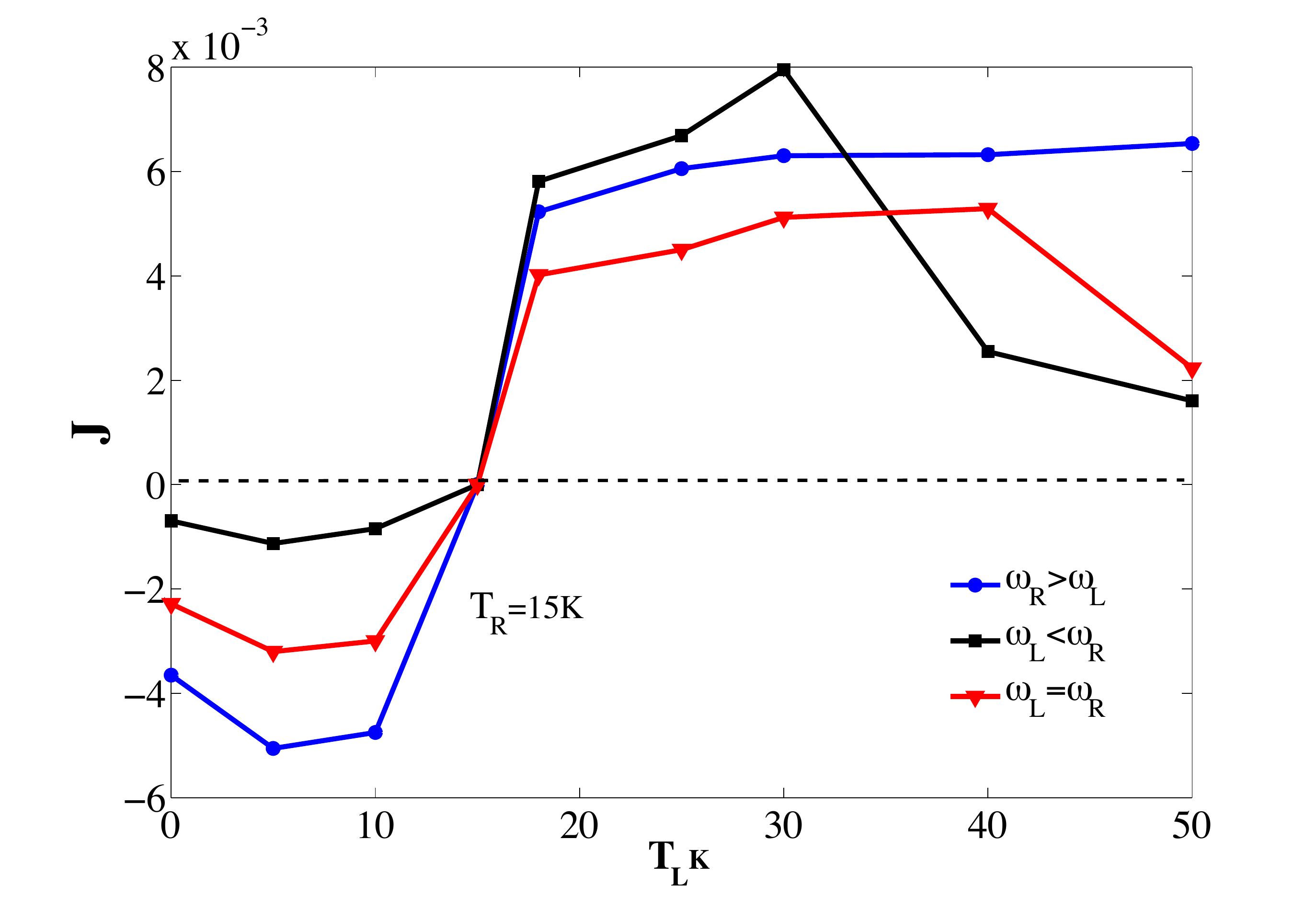}} 
\vspace{0.2cm} 
\caption{The heat current $J$ as a function of the temperature $T_L$ for fixed $T_R=15K$.
Three cases are shown blue $\omega_L > \omega_R$, black $\omega_L < \omega_R$
and red $\omega_L = \omega_R$.
} 
\label{fig:2}   
\end{figure}

Figure \ref{fig:2} shows the flux ${\cal J} $ as a function of $T_L$ for fixed $\omega_L$ , $\omega_R$ and $T_R$.
When $T_L$ becomes larger than $T_R$ the current switches direction in compliance with the II-law of thermodynamics.

For fixed right and left temperatures the steady state heat flux ${\cal J}$ is influenced by the frequency ratio $\omega_L/\omega_R$.
The heat flux ${\cal J}$ is larger when  $\frac{\omega_L}{\omega_R} > \frac{T_L}{T_R}$ compared to
$\frac{\omega_L}{\omega_R} < \frac{T_L}{T_R}$. This is an indication of the asymmetry of the model
and the rectifying effect.
\begin{figure}[tb]
\vspace{2.2cm} 
\center{\includegraphics[height=7.5cm]{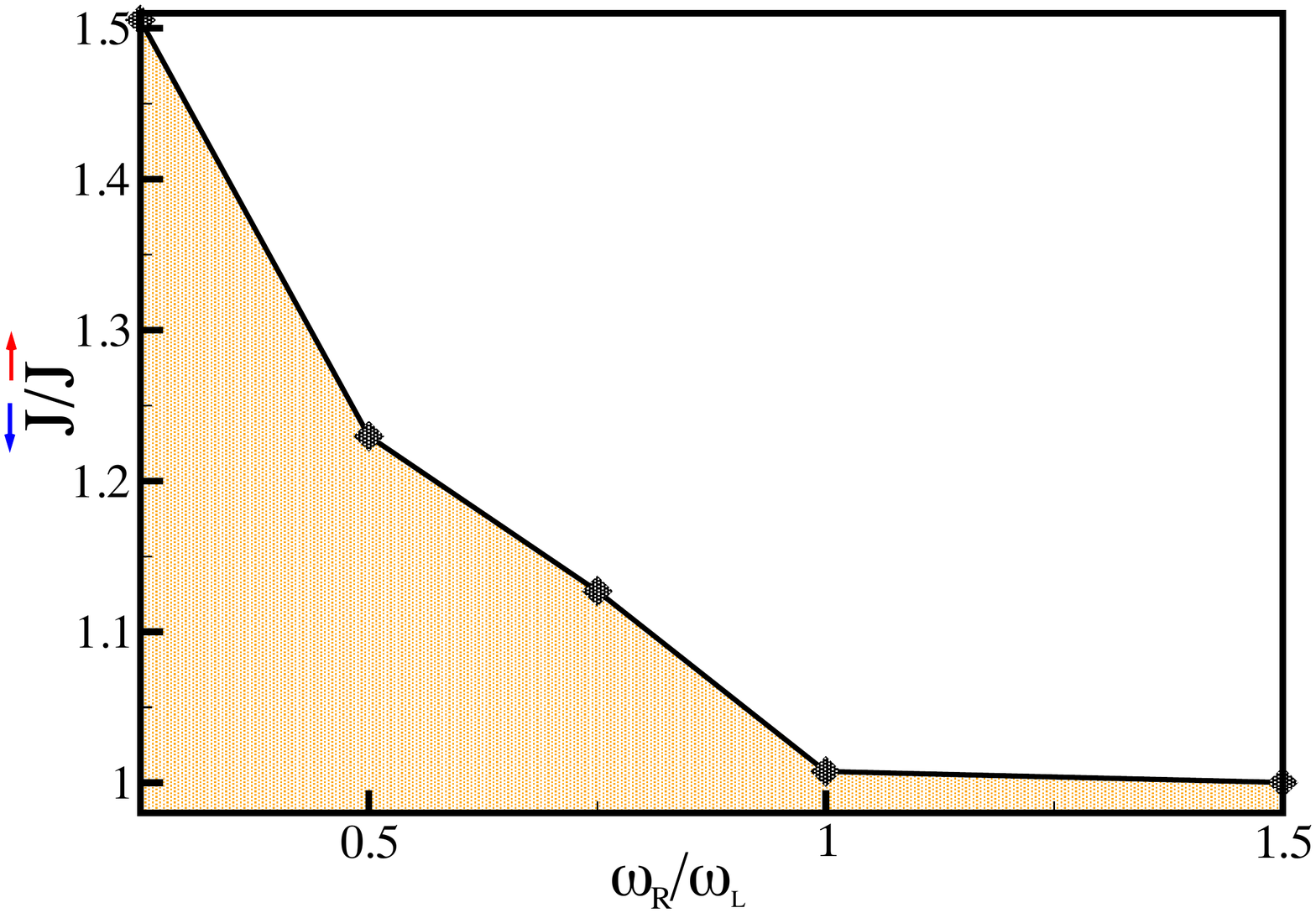}} 
\vspace{0.2cm} 
\caption{The rectifier effect: $\frac{ {\cal J}_{\rightarrow}}{{\cal J}_{\leftarrow}}$ as a function of the frequency  ratio $\omega_L/\omega_R$.
$\frac{ {\cal J}_{\rightarrow}}{{\cal J}_{\leftarrow}}$ is the ratio in the heat current 
from hot to cold  when the left and right baths are swithched. $T_c=5K$ and $T_h=25K$.
} 
\label{fig:2r}   
\end{figure}
The ratio between the hot to cold current $\frac{{\cal J}_{\rightarrow}}{{\cal J}_{\leftarrow}}$ when the left and right baths are exchanged leaving all other parameters 
fixed is shown in figure \ref{fig:2r} as a function of the frequency ratio $\frac{\omega_L}{\omega_R}$.
The product $\frac{{\cal J}_{\rightarrow}}{{\cal J}_{\leftarrow}} \frac{\omega_L}{\omega_R}$ is almost constant.

\begin{figure}[tb]
\vspace{2.2cm} 
\center{\includegraphics[height=7.5cm]{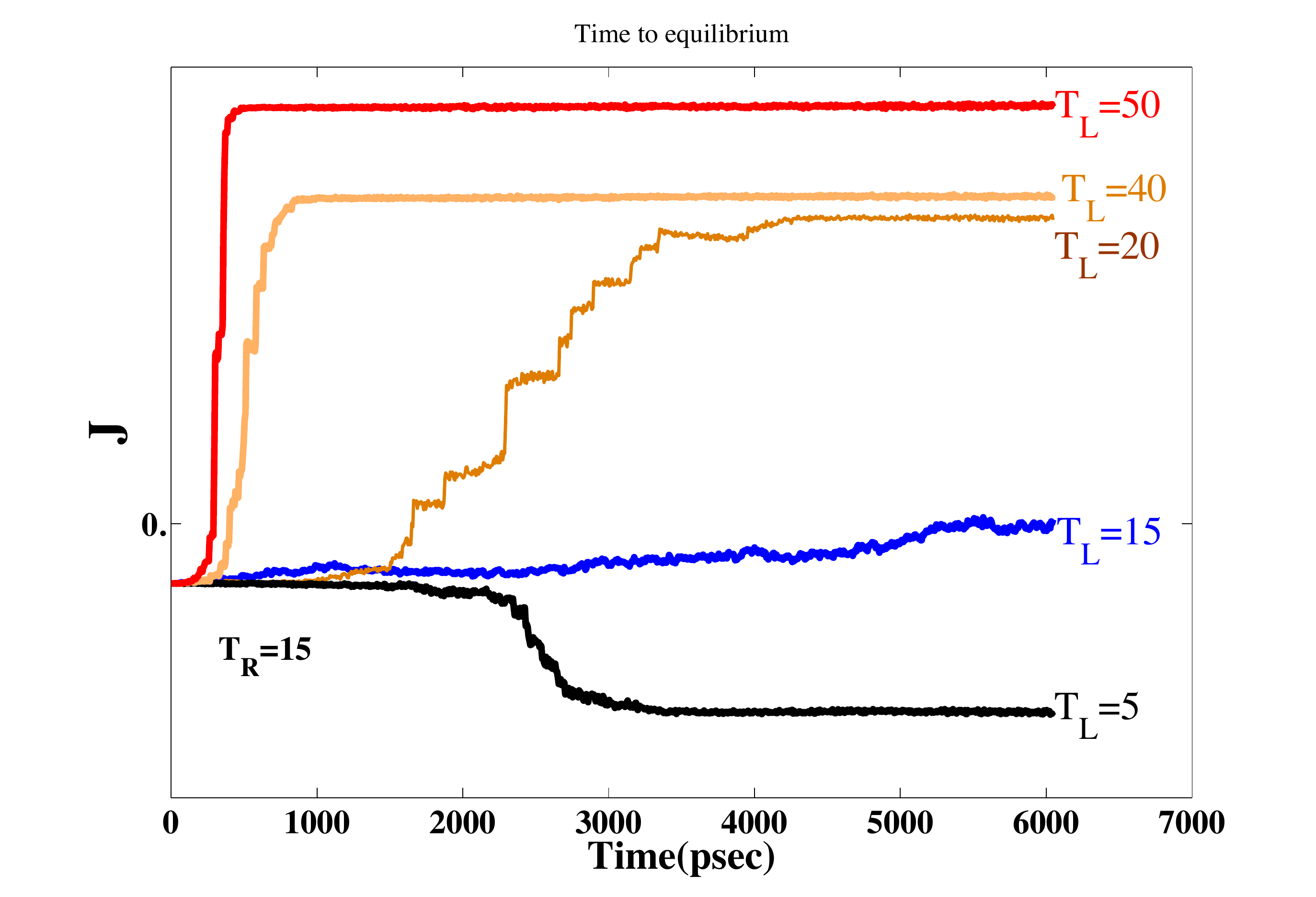}} 
\vspace{0.2cm} 
\caption{The Heat current $J$ as a function of time for different temperature   $T_L$ for fixed $T_R=15K$.
The timescale of approach to equilibrium depends on the temperature difference. The
number of realizations is 9.
} 
\label{fig:3}   
\end{figure}

An important dynamical effect is the rate of approach to steady state. Figure \ref{fig:3} shows the dynamics
of approach to steady state where the the right bath temperature $T_R$ is kept fixed and the left bath
temperature is varied. The relaxation rate is faster for larger temperature differences which would be expected from
Newtons heat law.

To conclude, the single molecular heat transport model complies with the first and second law of thermodynamics under all
conditions studied. 

\section{Heat pump operation.}
\label{sec:refrig}

A heat pump is a device that consumes power to drive heat from a cold to a hot bath.
We realize such a device by a single molecule connected to a hot and cold leads. The power is applied by a 
periodic electric field coupled to the molecule through the dipole operator.
The heat current through the system is calculated by accounting for the energy 
flux on each interface in addition to the energy flow from the external time dependent drive:
$$\Op H = \Op H_T + \Op V \cdot f(t)$$ 
where  $f(t) = \epsilon \cos(\nu t)$ and $\Op V= \mu \Op R$. 
We choose the driving frequency to be in resonance $\nu=\omega_L-\omega_R$.
$\epsilon $ is the driving amplitude and $\mu$ the dipole constant.

For the heat pump the energy currents become:
\begin{equation} 
 \frac{d E_S}{dt} = \langle i[\Op H_{SBR},\Op H_{S}]+i[\Op H_{SBL}, \Op H_{S}] \rangle +  \langle \frac{ \partial \Op H}{\partial t} \rangle
 ={\cal J}_R+{\cal J}_L + {\cal P}
\label{eq:cur}
\end{equation}  
This leads to: 
\begin{equation}
{\cal P} = \langle \Op V \dot f(t) \rangle = - \langle \Op R \rangle \epsilon \nu \sin (\nu t)
\end{equation}
In steady state the entropy production is generated only on the hot and cold baths:
\begin{equation} 
 \Delta {\cal S}_c+\Delta{\cal  S}_h = \frac{{\cal J}_h}{T_h} + \frac{{\cal J}_c}{T_c} \ge 0
\end{equation} 
The energy balance becomes ${\cal J}_h =- {\cal P}-{\cal J}_c$

\begin{figure}[tb]
\vspace{2.2cm} 
\center{\includegraphics[height=7.5cm]{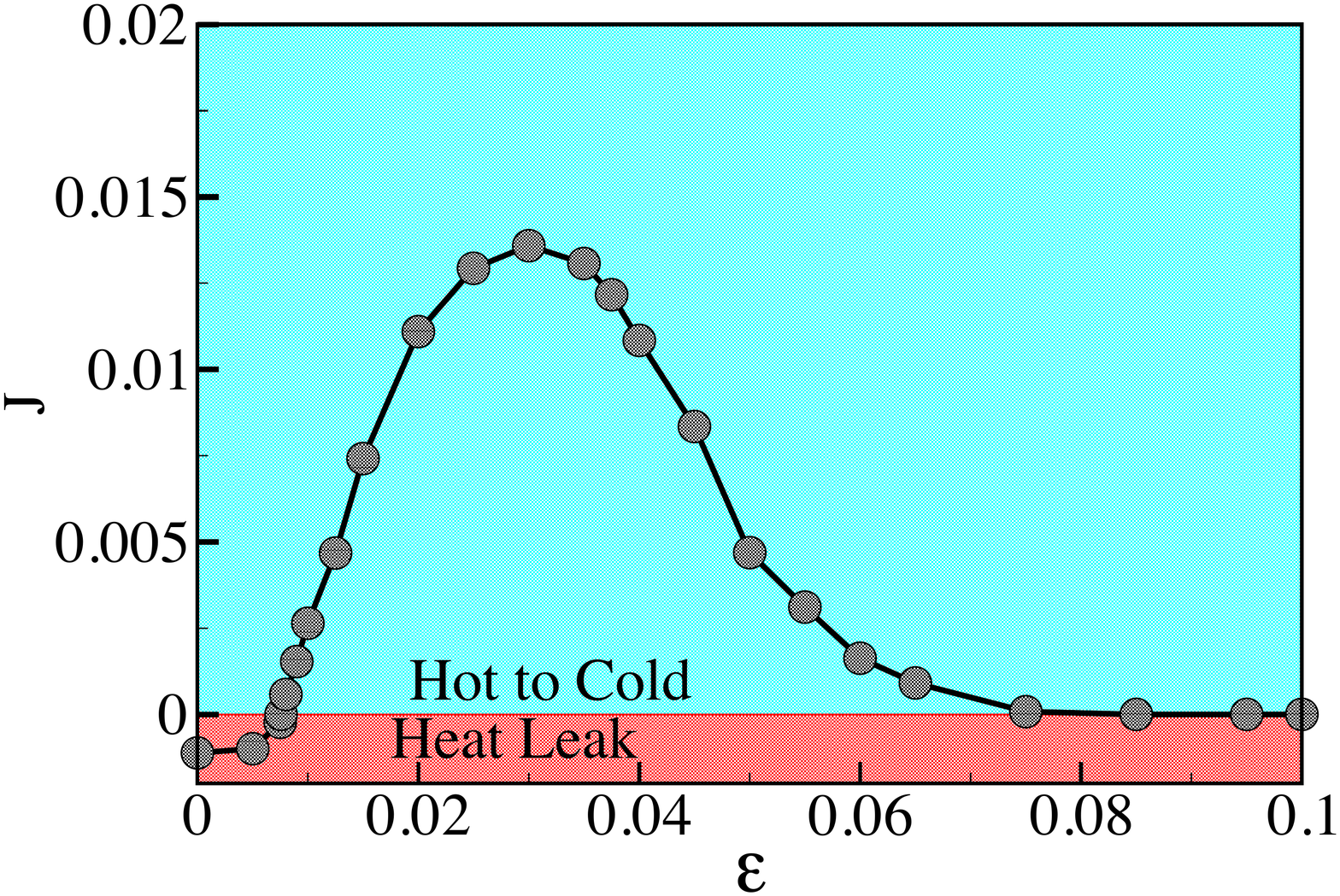}} 
\center{\includegraphics[height=7.5cm]{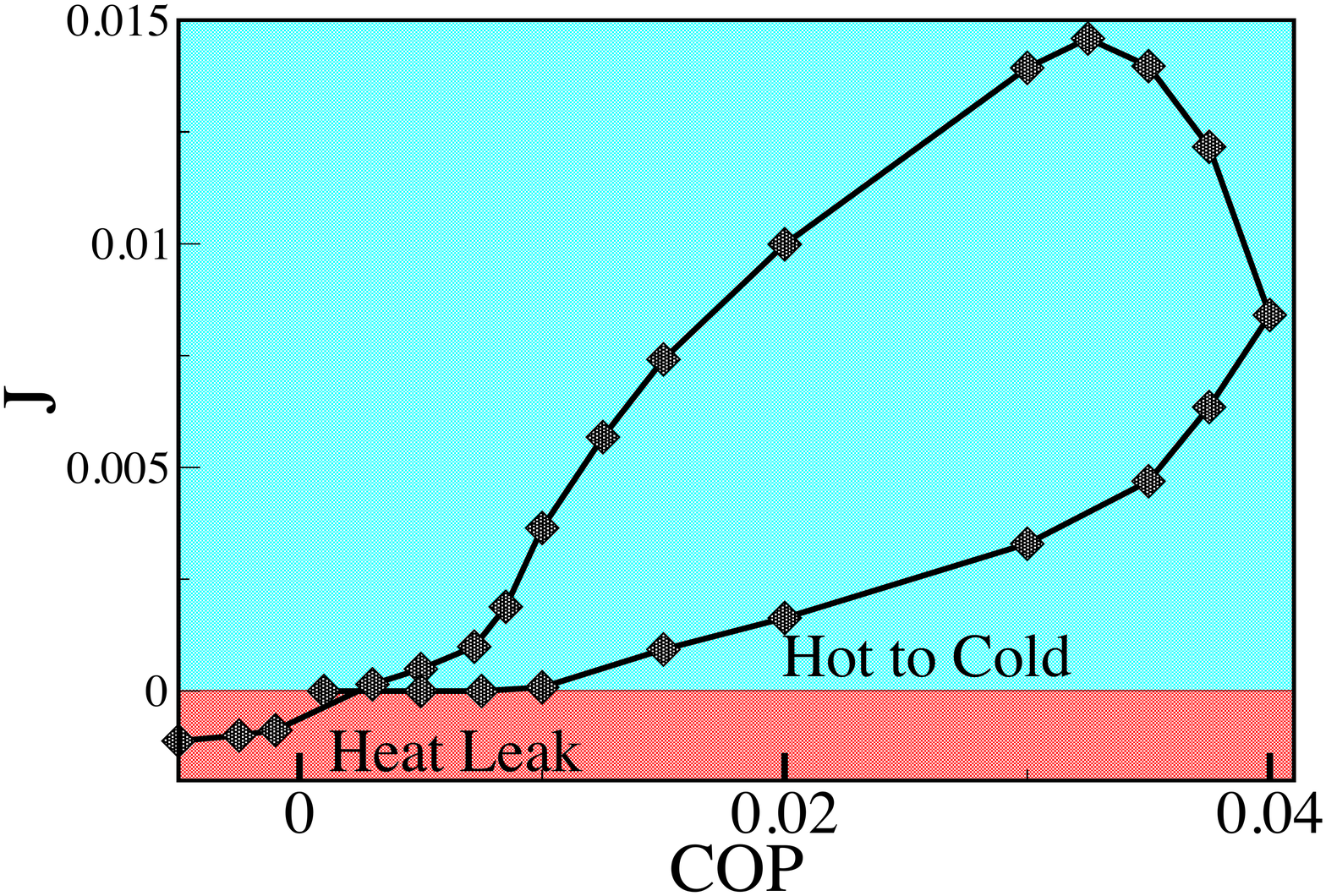}} 
\vspace{0.2cm} 
\caption{The heat pump: Top: Cooling current as a function of external driving amplitude  $\epsilon$.
Bottom: Cooling current as a function of the coefficient of performance (COP).
The bath temperatures are $T_c=10K$ and $T_h=25K$ and $\Gamma=0.5$.
A minimum threshold value of driving $\epsilon$ is required to overcome the heat leak (Red region).
Increasing the driving strength leads to a maximum in cooling which is due to a non-local system state.
The cooling decreases when the system localizes on the hot bath.
} 
\label{fig:4}   
\end{figure}

Figure \ref{fig:4} shows the cooling current ${\cal J}_c$ as a function of the amplitude $\epsilon$ of the driving field.
At weak driving  the heat that leaks from the hot to cold bath dominates. The driving is not strong enough to overcome the natural heat flow.
When the amplitude $\epsilon$  increases the heat flux ${\cal J}$ reverses sign pumping heat from the cold to hot baths.
At even stronger driving the derivative $\frac{ d {\cal J}_c}{d \epsilon}$ changes sign. The cooling current is reduced to zero.
The strong driving increases the systems energy and localizes the system on the hot bath. 

Figure \ref{fig:4} (bottom) shows the cooling current as a function of the COP, the coefficient of performance defined 
as $COP= \frac{{\cal J}_c}{\cal P}$. The loop in the graph shows two turning points of maximum cooling current and
maximum efficiency. This loop is typical of macroscopic classical refrigerators \cite{gordon1991generalized} and was
also observed in a model of a four-level quantum refrigerator \cite{correa2015internal}.

\begin{figure}[tb]
\vspace{2.2cm} 
\center{\includegraphics[height=7.5cm]{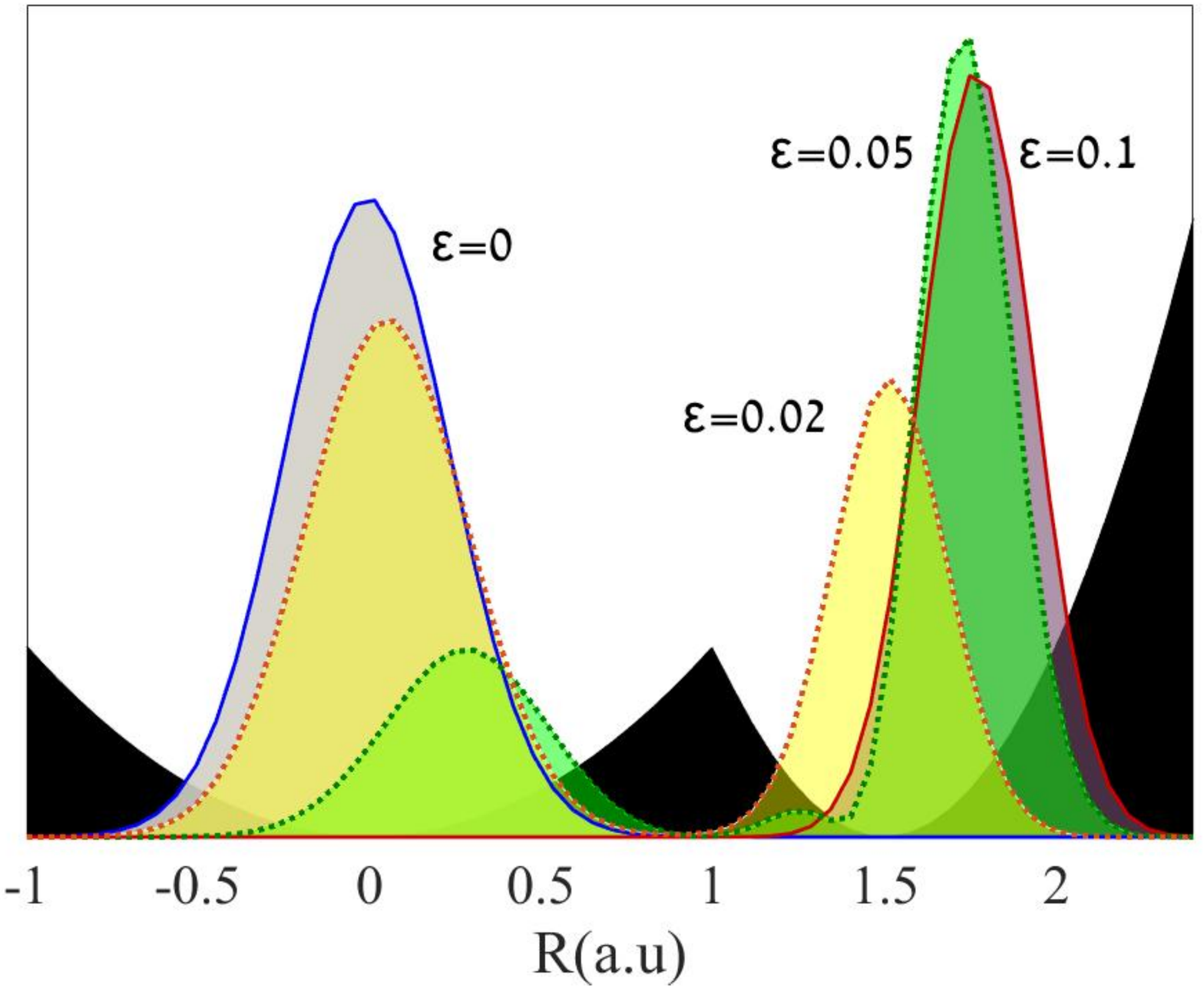}} 
\vspace{0.2cm} 
\caption{The probability density $p(R)=Tr \{ \hat \rho_s \Op R \}$  of the steady state of the system for different
driving amplitudes $\epsilon$ superimposed on the potential (black background).
Forsero driving the system localizes on the cold bath side. The optimum cooling is obtained for delocalized density $\epsilon=0.02$ and $\epsilon=0.05$.} 
\label{fig:5}   
\end{figure}

The change in the location of the steady state probability density is shown in Fig. \ref{fig:5} for different driving amplitudes $\epsilon$.
For small $\epsilon$ the density localizes on the cold potential well which is wider and has a lower zero point energy.
Increasing $\epsilon$ leads to delocalization accompanied by an increase in cooling power. Further increase in driving amplitude
localizes the density at the hot side with decreasing cooling power.

\begin{figure}[tb]
\vspace{2.2cm} 
\center{\includegraphics[height=7.5cm]{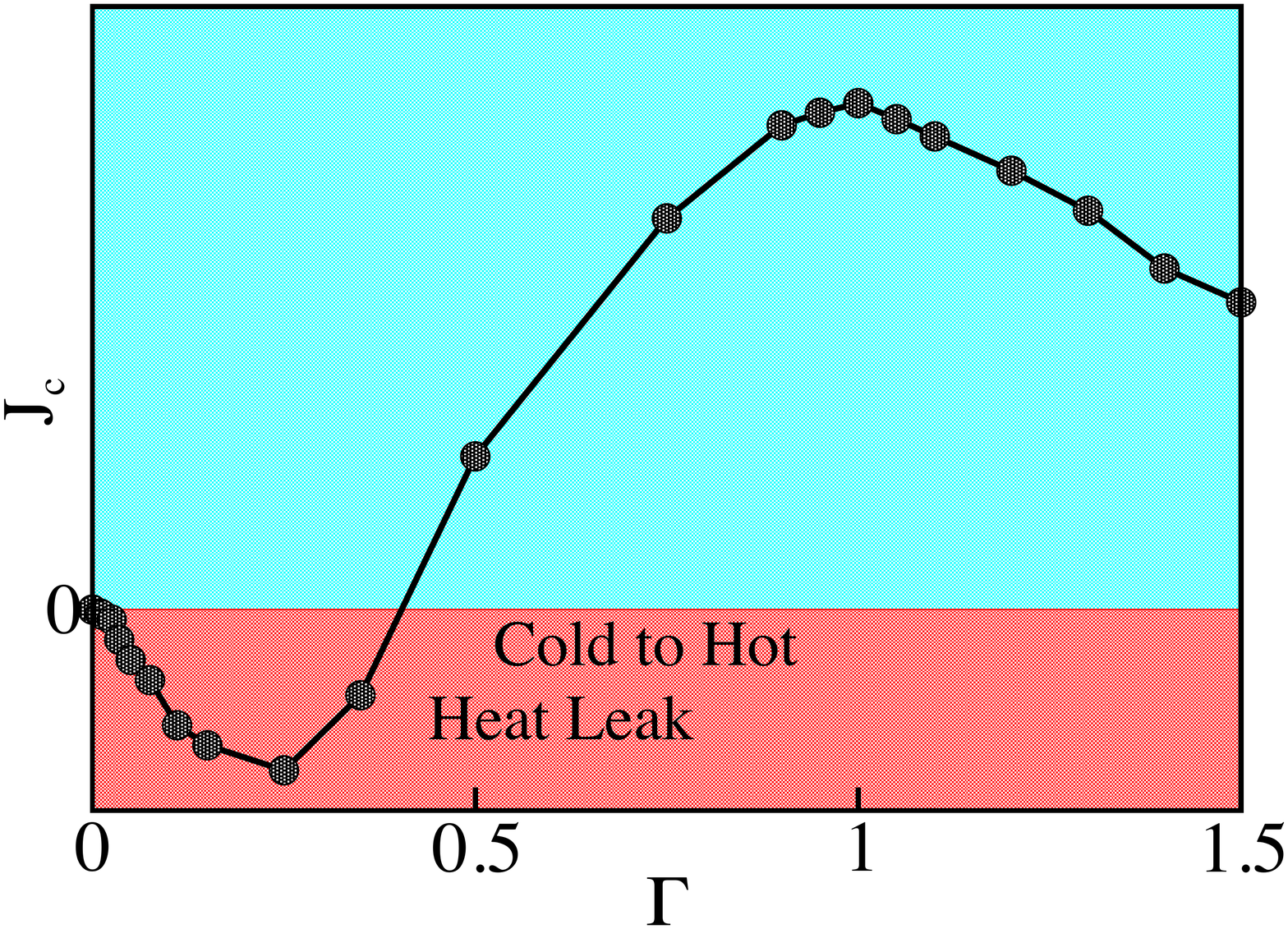}} 
\vspace{0.2cm} 
\caption{The cooling current as a function of the system bath coupling constant $\Gamma$.
Notice a maximum in dissipation at $\Gamma \sim 0.25$ and a maximum in coolling for $\Gamma \sim 1$.
The bath temperatures are $T_c=10K$ and $T_h=25K$ and the driving amplitude $\epsilon=0.25$ } 
\label{fig:6}   
\end{figure}

The transition form weak to strong coupling to the baths is demonstrated in Fig. \ref{fig:6}. When the coupling parameter
$\Gamma$ is smaller than the driving amplitude $\epsilon$ no refrigeration is observed. Analysis of the energy currents shows
that the heat current to the hot bath $J_h \approx 0$ therefore the external power is dissipated to the cold bath.
We find that the system's density is localized almost exclusively on the left well.
Increasing the coupling leads to a probability density in both walls. At this coupling value  there is a clear maximum in 
cooling. Further increase of the system-bath coupling  leads to a monotonic reduction in the cooling current. 
This is a manifestation of the strong coupling regime.
We find the probability density concentrating on the hot side leading to a loss of coherence which is essential for the refrigeration.

\begin{figure}[tb]
\vspace{2.2cm} 
\center{\includegraphics[height=7.5cm]{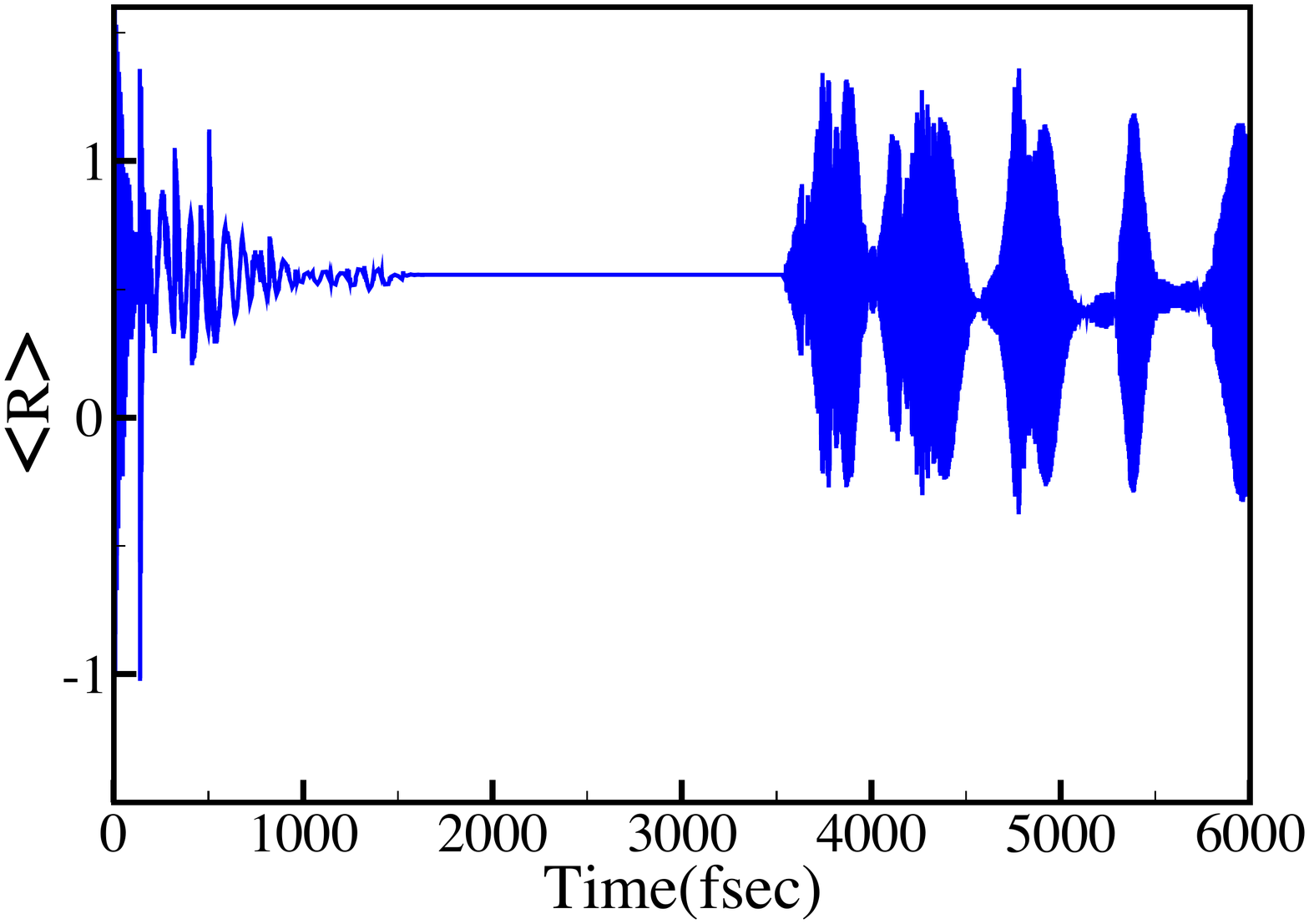}} 
\vspace{0.2cm} 
\caption{The expectation value of the position $\langle R \rangle$ as a function of time. The system is allowed
to relax to steady state reached after 1500 fsec. At $t=3500$ fsec, the coupling to the bath and the driving field is turned off. 
The  coherence in steady state manifests itself by complex dynamics. } 
\label{fig:7}   
\end{figure}

Continuous quantum devices typically contain coherence in order to perform their task of refrigeration \cite{k299,k306}.
Coherence exists when  the state of the system does not commute with its Hamiltonian. To demonstrate
this effect we first let the device reach steady state. This task was achieved in approximately 1500 fsec.
We then turned off all external couplings and followed the evolution in time. As a result the molecule 
undergoes periodic modulations. Figure \ref{fig:7} shows the dynamics of the mean position $\langle R \rangle$.
The initial dynamics is caused by the approach to steady state. The mean position is biased toward the right bath.
At the instant of 3500 fsec all connections are turned off ensuing periodic dynamics.

\section{Discussion}

We studied a quantum molecular device composed of a double well system coupled to a hot and a cold bath.
The model is based on a non-Markovian framework where the system and bath are strongly coupled.
The device operates as a heat rectifier facilitating heat transport when the high frequency part of the molecule is coupled to the hot bath.
Suppression of heat transport is obtained when the low frequency side is coupled to the hot bath.
When an external driving  in resonance with the frequency difference of the left and right wells is added then  the device operates as
a refrigerator pumping heat from the cold to the hot baths.
These features have been observed in properly constructed Markovian models based on the L-GKS equations. 
A proper construction of L-GKS equations which is conistant with the second law of thermodynamics requires a Floquet 
analysis and therefore is limited to periodic driving \cite{k122,k275,k278,k290,gelbwaser2013minimal,luis2016}. 
The equations of motion in the SSH approach are independent of the driving field, nevertheless maintaining 
compliance with the laws of thermodynamics.

The present model allows to gain insight on the regime of strong driving and strong system-bath coupling.
In a continuous device it is the coherence which is the enabler of operation \cite{k299,k306}. Direct evidence of coherence
is that the steady state reduced system state ${ \hat \rho_s}$ does not commute with $\Op H_s$.
This coherence is sensitive to the system-bath coupling parameter $\Gamma$. Over thermalisation causes the system
to localize in either the left or right potential wells. Optimal-performance is obtained from a delocalized global state.
This interplay explains the dependence of the cooling current ${\cal J}_c$ on $\Gamma$.

Refrigeration requires a minimum driving amplitude $\epsilon$ to overcome the intrinsic heat leak.
Overdriving leads to a decrease in ${\cal J}_c$. At low $\epsilon$ the system state is localized at the cold side
dissipating power to the cold bath. An increase in $\epsilon$ generates a non-local density supporting flux from the cold to hot bath.
Further increase in $\epsilon$ localizes the system on the hot side dissipating power to the hot bath.

The SSH method lends itself to the study of the role of fluctuations in the device currents. 
We can compare the current in a single  realization to the average over many realizations.
Due to the size of the Hilbert space in present study $\sim 10^7$ 
very few realizations are sufficient to converge the relevant observables.
We find that the number of realizations required  for convergence is smaller when steady state is reached.
A systematic study of fluctuation relations is still required.

To conclude, the stochastic surrogate Hamiltonian model is a viable tool in studying transport and driven transport
phenomena in the single device quantum regime. 

\subsection*{Aknowledgements}
We want to thank Amikam Levy, and Raam Uzdin  for their help , support and helpful discussions.
This work is supported by the Israel Science Foundation and 
by the European COST network MP1209.

%\subsection*{Bibliography}
% Style and layout of the references
%\bibliography{pub,transport-bib}

%\bibliographystyle{natbib}

%%MDPI internal note: new layout%% redefinition removed

%%%%%%%%%%%%%%%%%%%%%%%%%%%%%%%%

\end{document}